\documentstyle[12pt,epsfig]{article}
\newcommand{\woo}{\hskip -3pt\buildrel\circ\over\omega_0}
\def\ifig#1#2#3#4{\begin{figure}[hbtp]
 \begin{center}\leavevmode\epsfig{file=#1,height=#2}
 \caption[#3]{#4}\end{center}\end{figure}}

\title{Ground state properties of the $\beta$ stable nuclei in various
       mean field theories\footnote{This work is partly  
       supported by 
       the Polish Committee of Scientific Research under contract No. 
       2P3 03B 49 09}}

\author{                K.~Pomorski
        \thanks{On leave on absence from the University M.C.S. in Lublin}
        , P.~Ring, G.A.~Lalazissis\\
        Technische Universit\"at M\"unchen, Garching, Germany\\
                           and\\
        A.~Baran, Z.~\L ojewski, B.~Nerlo--Pomorska, M.~Warda\\ 
        Katedra Fizyki Teoretycznej, Uniwersytet M.C.S., Lublin, Poland}

\date{}

\newcommand{\beq}{\begin{equation}}
\newcommand{\eeq}{\end{equation}}
\newcommand{\bea}{\begin{eqnarray}}
\newcommand{\eea}{\end{eqnarray}}
\newcommand{\eql}[1]{\label{eq:#1}}
\newcommand{\req}[1]{(\ref{eq:#1})}
\newcommand{\larr}{\leftarrow}
\newcommand{\rarr}{\rightarrow}

\begin{document}

\maketitle
          
\begin{abstract}
The separation energies of neutrons and protons, binding energies, 
mean square charge
radii, electric quadrupole moments and deformation parameters of the
proton and neutron distributions are evaluated for  $\beta$ stable
even--even nuclei with $16 \leq A \leq 256$. We compare the theoretical
estimates obtained within the Hartree-Fock plus BCS model with a few
sets of Skyrme forces, relativistic mean--field theory and frequently
used Saxon-Woods and Nilsson potentials with experimental data.
\end{abstract}

\section{Introduction}

Mean field theory is a powerful tool for the description of
low energy nuclear phenomena.  Effective microscopic
theories have been developed, the most fruitful being the
Hartee-Fock calculations with effective density dependent
interactions. Forces of Skyrme type \cite{Sk56,Va72} with zero range 
or of Gogny \cite{Go75,Be84} type with finite range are among the most
successful. In recent years a relativistic description of
the ground state properties of nuclei has also been
proposed. Relativistic mean field (RMF)  theory has been
quite successful in describing finite nuclei at $\beta$
stability line as well as far away from it \cite{SW86}.

Apart from the microscopic effective theories, the
macroscopic-microscopic models have also been developed.
The Nilsson \cite{Ni57} and the Saxon-Woods
\cite{Wahl} potentials have
been widely used providing satisfactory results.
    
For the effective theories and the macroscopic-microscopic
shell correction models mentioned above several parameter
sets  have been proposed. The number of the parameters
varies in the different approaches. They are usually
determined by a global fit to various ground state nuclear
properties of $\beta$ stable nuclei. The scope is always a
better description of the available data and a hope to
achieve in this way a higher predictive power for nuclei
far away from stability.

The aim of this work is to analyze the results of several
effective mean-field theories and of the shell correction models
for the description of separation energies, sizes and shapes of
neutron and proton distributions in the ground state of the
$\beta$--stable even-even nuclei in a wide range of the mass
numbers. The results are obtained using the most recent or
frequently used parametrizations for each theory or model. This
allows a systematic comparison of the different results
of each theory close to $\beta$--stability. A reliable
description of the nuclear ground state properties along the
$\beta$--stability line is essential for a successful
extrapolation to exotic nuclei as well as to superheavy nuclei.

In this paper we choose to investigate the Skyrme \cite{Sk56,Va72} mean
field theory and the RMF \cite{SW86} theory together with average
potential models of Nilsson \cite{Ni57,Gu67} and
Saxon-Woods \cite{Cw87} type. The calculated quantities
are the total binding energies ($B$), the proton
($S_p$) and neutron ($S_n$) separation energies, the mean
square charge radii (MSR) and electric quadrupole moments
($Q_2$). The equilibrium deformations and root mean square
radii (RMS) of proton and neutron distributions are also
evaluated.

In Section 2 a brief discussion of the formalism used in
each approach is presented. In Section 3 we give some
details about the calculations and the observables under
consideration. In Section 4 the results are presented and
discussed. Finally in Section 5 the main conclusions are
summarized.

\section{The theoretical formalism}

For a long time many results have been obtained by
the macroscopic-microscopic model using phenomenological
single-particle potentials. In the present investigation
we have employed Nilsson and Saxon-Woods potentials. A
short presentation of these models is given in Subsections 2.1 and
2.2. 

Hartree-Fock approach based upon phenomenological density
dependent zero range forces of Skyrme type has proved to be very
successful in the microscopic description of ground state
properties of nuclear matter and of finite nuclei over the entire
periodic table. The Skyrme forces are presented in
Subsection 2.3. In recent years relativistic mean field 
theory with nonlinear self-interactions between the mesons has
gained in considerably interest for the investigations of
low-energy phenomena in nuclear structure. With only a few
phenomenological parameters such theories are able to give a
quantitative description of ground state properties of spherical
and deformed nuclei at the stability line and far away from it.
In RMF theory the nucleons are described in a fully relativistic
way as Dirac spinors ineracting via the exchange of various
mesons. In that sense the theory appears to be more fundamental
since both nucleonic and mesonic degrees of freedom are taken
into account.  Moreover, the spin-orbit interaction is treated
correctly and no extra parameter is necessary. In the following
(Subsection 2.4) we give a short description of the formalism of
these theoretical approaches.

\subsection{The single particle potential of Nilsson}

The modified anisotropic harmonic oscillator potential was
introduced by Nilsson in Ref. \cite{Ni57} for the
quadrupole deformation ($\varepsilon$) and a few years
later it was generalized in Ref. \cite{Gu67} for
deformations of higher multipolarity
($\varepsilon_\lambda$). The single particle hamiltonian
with the Nilsson potential has the following form in the
stretched coordinates system
\begin{eqnarray}
\hat H_{sp} &=& \frac{1}{2} \hbar\omega_0(\varepsilon,\varepsilon_4, ...)
 ( -\Delta_\rho + \frac{1}{3}\varepsilon(2\frac{\partial^2}
 {\partial\zeta^2} - \frac{\partial^2}{\partial\xi^2} -
 \frac{\partial^2}{\partial\eta^2}) \nonumber \\ 
&+&  \rho^2\left [1 - \frac{2}{3}\varepsilon P_2(cos\theta_t) + 2\varepsilon_4
     P_4(cos\theta_t) +  ....\right ] )  + V_{corr}.
\end{eqnarray}
where $\rho^2=\xi^2+\eta^2+\zeta^2$ and the dimensionless
stretched coordinates are defined as follows:
\beq 
\xi=\sqrt{\frac{M\omega_\perp}{\hbar}}x \;\;, \;\;
\eta=\sqrt{\frac{M\omega_\perp}{\hbar}}y \;\;, \;\;
\zeta=\sqrt{\frac{M\omega_z}{\hbar}}z \;\;, 
\eeq
Here
\beq
\omega_\perp = \omega_0(\varepsilon,\varepsilon_4, ...) 
(1+\frac{1}{3}\varepsilon)\;\;\;\;  {\rm and}\;\;\;\;
\omega_z = \omega_0(\varepsilon,\varepsilon_4, ...) 
(1-\frac{2}{3}\varepsilon) \;\;.
\eeq
The deformation dependence of the harmonic oscillator
frequency $\omega_0$ is obtained from the volume
conservation condition
\beq
\frac{\omega_0^3}{\woo^3} = \frac{1}{(1+\frac{1}{3}\varepsilon)
 (1-\frac{2}{3}\varepsilon)^{1/2}} \int_0^1 \frac{d (cos \theta_t)}
  {\left [1 - \frac{2}{3}\varepsilon P_2(cos\theta_t) + 2\varepsilon_4
 P_4(cos\theta_t) +  ....\right ]^{3/2}} \;\; .
\eeq
The frequency $\hbar\woo$ of the spherical harmonic
oscillator was taken from Ref. \cite{Ne93}
\beq
\hbar\woo = 40 / A^\frac{1}{3} MeV \;\; .
\eeq
The correction potential $V_{corr}$ in the Nilsson
hamiltonian contains the spin--orbit ($V_{ls}$) and $l^2$
($V_{l^2}$) terms
\beq
V_{corr} = V_{ls} + V_{l^2} \;\;\; ,
\eql{vcorr}
\eeq
which were taken in the form proposed in Ref. \cite{Se86}:
\beq
V_{ls} = -2\hbar\woo \kappa_{Nl}\; \vec{l}\cdot\vec{s} \;\;\;,
\eeq
\beq
V_{l^2}=-\hbar\woo\left[\nu_{Nl}\;\vec{l}^2-<\nu_{Nl}\;
\vec{l}^2>_N\right]\;\;\;.
\eeq
The coeficients $\kappa_{Nl}$ and $\nu_{Nl}$ are evaluated
from the following expressions:
\beq
\kappa_{Nl} = \kappa_0\left [1+8\;\nu_{Nl}(N+\frac{3}{2})\right ] +
\kappa_1 A^{-1/3} \int_{R_0-a/2}^{R_0+a/2} {\cal R}^2_{Nl}\;r^2
dr\;\;\;, 
\eeq
\beq
\nu_{Nl} = \nu_0 \left[\int_0^{R_0-a/2} {\cal R}^2_{Nl}\;r^2
dr\right]^2\;\;\;, 
\eeq
where $R_0=1.2 A^{1/3} fm$ is nuclear radius, $a=0.7 fm$ is
the surface thickness and ${\cal R}_{Nl}(r)$ is the harmonic
oscillator radial wave function.  The correction term
\req{vcorr} in the Nilsson potential depends on three
adjustable parameters only \cite{Se86}: 
\beq
\kappa_0=0.021,
\;\;\;\; \kappa_1=0.90 \;\;\; and \;\;\; \nu_0=0.062\;\;,
\eeq
which are valid in the all mass regions.

The potential energy of nucleus is calculated by the shell correction method
\cite{St66} with the final range liquid drop (FRLD) macroscopic term 
\cite{Mo95} which contains also several fenomenological
parameters. 

\subsection{The single-particle potential of Saxon-Woods}

The deformed Saxon-Woods potential is widely described in
the literature \cite{Cw87} and we restrict ourselves to
represent only the basic formulae. The potential consists
of the central part $V_{cent}$, the spin-orbit part
$V_{so}$ and the Coulomb potential $V_{Coul}$ for the
protons:
\beq
V^{WS}(\vec{r},\vec{p},\vec{s};\beta)= V_{cent}(\vec{r};\beta)+
V_{so}(\vec{r},\vec{p},\vec{s};\beta)+V_{Coul}(\vec{r};\beta) 
\eql{ws}
\eeq
with
\beq
V_{so}(\vec{r},\vec{p},\vec{s};\beta)=
-\lambda (\nabla V_{cent}\times\vec{p})\cdot\vec{s}. 
\eql{vso}
\eeq
The central part is defined by:
\beq
V_{cent}(\vec{r};\beta) =\frac{V_0[1 \stackrel{+} - \kappa(N-Z)/A]}
{[1+exp(l(\vec{r};\beta)/a)]},
\eql{vc}
\eeq
where $a$ is the diffuseness of the nuclear surface. The
set of deformation parameters $\beta_{\lambda}$, which
characterize the nuclear shape, is denoted by $\beta$. The
function $l(\vec{r},\beta)$, describing the distance
between the given point $\vec{r}$ and the nuclear surface
has been determined numerically \cite{Cw87}. For spherical
nuclei we have $l(\vec{r},\beta=0)\;=\;r-R_0$, where
$R_0=r_0 A^{1/3}$, is the radius of the corresponding
spherical nucleus.
\beq
R(\theta) = c(\beta)R_0[1+\sum_{\lambda} \beta_{\lambda}Y_{\lambda 0}
(cos(\theta )] \;\;\; .
\eql{rbeta}
\eeq
The function $c(\beta)$ insures the conservation of the nuclear 
volume with a change of the nuclear shape.

The various sets of Saxon-Woods potential parameters are
presented in Table 1. The "universal" \cite{Cw87}, Wahlborn
\cite{Wahl}, Rost \cite{Rost}, Chepurnov \cite{Chep} and the
"new" \cite{New}  ones were chosen.

\subsection{Effective mean field theory with Skyrme forces}

Mean field theory of the Skyrme type starts from an energy
functional, which is derived from a density dependent
two-body interaction of the form:
\newpage
\bea
 V_{12} 
   &=& t_0 (1 + x_0 P_\sigma)\,\delta (\vec r_1-\vec r_2)
         \nonumber \\
   &-& \frac{1}{2} t_1 (1 +x_1 P_\sigma) 
       \left [{\stackrel\larr\nabla}_{12}^2 \,\delta (\vec r_1-\vec r_2) + 
       \delta (\vec r_1-\vec r_2) {\stackrel\rarr\nabla}_{12}^2 \right ] 
       \nonumber \\
   &-& t_2\,(1 + x_2 P_\sigma) {\stackrel\larr\nabla}_{12} \,
       \delta (\vec r_1-\vec r_2){\stackrel\rarr\nabla}_{12}
         \\
   &+& \frac{1}{6} t_3 (1 + x_3 P_\sigma )
       \left [\rho_{q_1}(\vec r_1) + \rho_{q_2}(\vec r_2)\right ]^{\gamma}
       \delta (\vec r_1-\vec r_2)         \nonumber \\
   &-& i \omega_0 {\stackrel\larr\nabla} _{12} \wedge \delta (\vec r_1-\vec r_2) 
       {\stackrel\rarr\nabla}_{12}\cdot(\vec\sigma_1 + \vec\sigma_2) 
       + V_{Coul}\,\,,
         \nonumber
\eql{Skyrme}
\eea
where $t_0, t_1, t_2, t_3, x_1, x_2,
x_3, w_0$ and $\gamma$ parameters are shown in Table 2 as 10
sets called: $i$ \cite{Va72}, $ii$ \cite{Va73}, $iii$, $iv$, $v$, $vi$
\cite{Be75}, $vii$ \cite{Gi80}, $m^\star$ \cite{Ba82}, $a$ \cite{Ko76}, 
$p$ \cite{Do83}. $P_\sigma$ is the spin exchange operator.

Using Slater determinants as variational functions the
energy functional of a nucleus can be expressed 
as a volume integral: 
\beq
 E = \int H(\vec r) d^3\vec r \,\, ,
\eql{ESk}
\eeq
where the energy density $H(\vec r)$ is a function of
the nucleon density $\rho$, the kinetic energy density
$\tau$ and spin density $\vec J$. The total energy
\req{ESk} has to be minimized with respect to the choice of
the many-body wave functions, which leads to a non-linear
eigenproblem with the corresponding mean-field hamiltonian.

\subsection{The relativistic Hartree formalism}

RMF theory is based on an effective lagrangian containing
both nucleonic and mesonic degrees of freedom
\cite{SW86,PR96}:
\bea
{\cal L}&=& \bar\psi_i\{i\,\gamma^\mu\,\partial_\mu - M\}\psi_i \nonumber \\
  &&+\frac{1}{2}\partial^\mu\sigma\,\partial_\mu\sigma 
    -U(\sigma) -g_\sigma\,\bar\psi_i\psi_i\,\sigma \nonumber \\
  &&-\frac{1}{4}\Omega^{\mu\nu}\Omega_{\mu\nu} 
    +\frac{1}{2}m_\omega^2\,\omega^\mu\omega_\mu 
    -g_\omega\,\bar\psi_i\,\gamma^\mu\,\psi_i\,\omega_\mu \\
  &&-\frac{1}{4}\vec{R}^{\mu\nu}\vec{R}_{\mu\nu} 
    +\frac{1}{2}m_\rho^2\,\vec\rho\,^\mu\vec\rho_\mu 
    -g_\rho\bar\psi_i\,\gamma^\mu\vec\tau\,\psi_i\,\vec\rho_\mu  \nonumber \\
  &&-\frac{1}{4}F^{\mu\nu}F_{\mu\nu} 
    -e\bar\psi_i\,\gamma^\mu\,\frac{(1-\tau_3)}{2}\,\psi_i\,A_\mu\;, \nonumber
\eql{lrmf}
\eea
where $M$ is the nucleon mass, $m_\sigma, m_\omega, m_\rho$
are meson masses, and $g_\sigma, g_\omega, g_\rho$ are meson
coupling constants. The isovector quantities are indicated
by arrow bars. In our approach we use the non-linear
$\sigma$-$\omega$-$\rho$ model, that is we consider that the sigma
mesons move in their non linear potential created by the other ones:
\beq
U(\sigma)=\frac{1}{2} m^2_\sigma\sigma^2
  +\frac{1}{3}g_2\sigma^3+\frac{1}{4}g_3\sigma^4 \,\,.
\eeq
The parameters of RMF theory are gathered in Table 3:
 NL-1 \cite{Re86}, NL-3 \cite{La96}, NL-SH \cite{Sh93}. 

The classical variational principle yields a coupled set of
equations of motion for the Dirac spinors and the fields:
the Dirac equation for the nucleons, Klein Gordon equations
for the meson fields and the Laplace equation for the
Coulomb field.  The solution of this non-linear set of
coupled equations is carried out iteratively. 
The calculations for both the Skyrme mean field as well as
the RMF theory have been carried out in the axial symmetric
configuration using the oscillator expansion method
\cite{Ga90}.

\section{Details of the calculations}

Since most of the nuclei considered here are open shell
nuclei, pairing has been included using the BCS formalism.
We have used constant pairing gaps for protons and neutrons
which have been obtained from the empirical particle
separation energies by the formulae:
\beq
\Delta_p(Z,N) = \frac{1}{4} \left( B(Z-2,N) - 3B(Z-1,N) + 3B(Z,N) - B(Z+1,N) 
                \right)\;\;,
\eql{dp}
\eeq
\beq
\Delta_n(Z,N) = \frac{1}{4} \left( B(Z,N-2) - 3B(Z,N-1) + 3B(Z,N) - B(Z,N+1) 
                \right)\;\;.
\eql{dn}
\eeq

It should be noted that for constant pairing gaps $\Delta_{n(p)}$
the pairing energy diverges if it is extended over an
infinite configuration space. Therefore in all of our
calculations a pairing window has been considered.

We have chosen nuclei with the smallest mass for a given
nucleon number A. Obviously these nuclei are stable against
$\beta$--decay. We have considered in our analysis the
even--even nuclei with $16 \leq A \leq 256$.  For all these nuclei
calculations for the total binding energy have been carried
out.

For an estimate of the proton (neutron) separation energies
an approximate method is proposed.  It is known  from the
BCS theory that in order to separate a nucleon from an
even-even nucleus one has to break the Cooper pair and to
move this nucleon from the Fermi level ($\lambda$) to the
continuum limit. This experimental relation can be written as
follows:
\beq
\lambda_p(Z,N) = \Delta_p(Z,N) - S_p(Z,N) \;\;\ ,
\eql{lp}
\eeq
\beq
\lambda_n(Z,N) = \Delta_n(Z,N) - S_n(Z,N) \;\; .
\eql{ln}
\eeq

The values of these 'experimental'  Fermi energies ({\it
exp}, diamonds) are plotted in Figs. 1 and 2 for protons
and neutrons respectively.  The solid lines ({\it avr}) in
Figs. 1 and 2 correspond to the average position of the
Fermi level obtained by the phenomenological formula:
\beq
\bar\lambda_p = \frac{-11.8\, MeV}{1+ A/208} \;\; ,
\eql{avlp}
\eeq
\beq
\bar\lambda_n = \bar\lambda_p - 0.512\, MeV \;\; .
\eql{avln}
\eeq
The coefficients in Eqs. \req{avlp} and \req{avln} were
obtained by the least square fit to the experimental values
of the Fermi energies of all $\beta$-stable nuclei. The
average position of the Fermi level for neutrons is shifted
down with respect to the Fermi level for protons by one
electron mass as one would expect from general
thermodynamical considerations.

The values of $\bar\lambda_p$ and $\bar\lambda_n$  \cite{Au93} as well as
the average MSR radii of nuclei \cite{Ne94} can be used to
establish the depth and size of the average potential well for
$\beta$--stable nuclei. 

We are interested in quantities characterizing the neutron
and proton distribution in nuclei, their sizes and shapes.
Thus, we have performed systematic calculations of the
proton and neutron mean square radii and of the proton and
neutron deformation parameters $\beta_{n(p})$.
  
The charge radii were calculated from the corresponding
proton radii taking into account the correction due to
finite proton size:

\beq
<\vec r\,^2>_{ch} = <\vec r\,^2>_p + 0.64\,\,fm^2 .
\eeq

We have neglected here the small contributions to the mean charge
square radius originating from the electric neutron form
factor and the electromagnetic spin-orbit coupling
\cite{Be72,Ni87}.

The global measure of the deformation of the neutron (or
proton) distribution in the case of the microscopic
theories can be expressed by the corresponding quadrupole
moment
\beq
 <Q_2>_{n,p} = <2r^2 P_2(cos\theta)> \,\,\,.
\eeq
Having the quadrupole and monopole moments we can estimate
approximately the qua\-dru\-po\-le deformation parameter $\beta$
of the neutron (or proton) distribution \cite{Ha88}
\beq
\beta_{n,p} = \sqrt{\frac{\pi}{5}}\frac{<Q_2>_{n,p}}{<Q_0>_{n,p}} \,,
\eql{beta}
\eeq
where $<Q_0>_n = N<\vec r\,^2>_n$ and $<Q_0>_p = Z<\vec
r\,^2>_p$.  This simple estimate for the quadrupole
deformation is valid only for small deformation parameters 
$\beta$.

The reduced electric quadrupole transition between the
rotational $2^+$ and $0^+$ states are proportional to the
square of the proton quadrupole moment 
\beq
B(E2) = \frac{5}{4 \pi}<Q_2>^2_p \;\; .
\eeq

\section {Numerical results and discussion} 

\subsection{Macroscopic-microscopic models}

Using the single-particle Nilsson potential and the well
known Strutinsky \cite{St66} shell correction method with
the renormalisation to the finite range liquid droplet model
\cite{Mo95} we have found the ground state deformations of
all the investigated nuclei. We are not going to compare
here the total binding energy of a nucleus with the
experimental data because this observable is not only
dependent on the potential but also on the macroscopic
part. We know from the extended calculations made e.g. in
Ref. \cite{Mo95} that the advanced macroscopic-microscopic
model leads to very good estimates of the nuclear masses
much closer to the experimental data than those obtained
within the microscopic theories. One should keep in mind,
however, the large number of parameters of these models as
well as the large experimental input used for their
determination.

Having the ground state deformations we have evaluated
microscopically the mean square charge radii (MSR) and
the reduced electric quadrupole transition probabilities
($B(E2)$) in the corresponding equilibrium points.  The
Nilsson model estimates of the radii $r_{ch}$ are plotted
in Fig. 3 and of $B(E2)$-values in Fig. 4. The theoretical
results are compared with the experimental data taken from
Refs. \cite{Vr87,Ot89}. It is seen in Fig. 3 that the Nilsson
average potential reproduces very well the experimental
trend of $r_{ch}$ in the whole range of A. The mean square
deviation of the theoretical values from the experimental
data is
\beq
 <{\Delta}^2 r_{ch}>^{1/2} = 0.029\, fm .
\eeq
The calculated $B(E2)$ transition probabilities are also
close to the experimental values:
\beq
 <{\Delta}^2 B(E2)>^{1/2} = 1.336 e^2b^2 \;\;.
\eeq
Only in the region of $A \approx 220$ are our estimates of
$B(E2)$ too large. It is probably due to the lack of
octupole deformation of these heavy nuclei in our analysis.
We have performed the calculations with the Nilsson
potential for one set of the correction term parameters
\cite{Se86}  valid in the whole range of $A$ \cite{Ma95}.

The infinite Nilsson potential is not adjusted for
reproducing the separation energies of the neutron ($S_n$)
or of the proton ($S_p$). We can easily obtain such
estimates with the Saxon-Woods potential and we have
performed an extensive calculation for five commonly used
sets of parameters (see Table 1). The results are presented
in Fig. 5, where the differences between the
theoretical and the experimental \cite{Au93} separation
energies of protons and neutrons are plotted.  The best
theoretical estimates of $S_n$ and $S_p$ were obtained with
the Saxon-Woods potential of Chepurnov \cite{Chep}. For the sake
of comparison we present also the results 
for the commonly used  'universal' set of
parameters \cite{Cw87} (Fig. 5).

The mean square deviations between the theoretical and
experimental values of $S_n$, $S_p$, $B(E2)$ and $r_{ch}$
for all investigated Saxon-Woods models are plotted in Fig.
6. It is seen that  the Chepurnov set apart from the charge
radii gives the smallest deviations from the experiment.  For
the charge radii $r_{ch}$  the other sets are somewhat
better than Chepurnov parametrization.

\subsection{Self-consistent theories}

A similar analysis of the $\beta$-stable even-even nuclei
features has also been performed within the Hartree-Fock
model with density dependent Skyrme forces as well as
within the RMF theory.

The root mean square errors for the neutron ($S_n$) and
proton ($S_p$) separation energies, the binding energies
per particle ($B$), the reduced transition probabilities
($B(E2)$) and charge radii ($r_{ch}$) were obtained within
the HF+BCS scheme for ten different sets of the Skyrme
forces (see Table 2). These mean square deviations are
plotted in Fig. 7. It is seen that on the average the
predictions of $m^*$ and $p$ effective forces are closest
to the experiment. It is noted however, that the other sets
give also quite reasonable predictions for certain
observables.

We have also investigated the deviation of the
calculated quantities  from those of the empirical
values as a function of the mass number. This is given in
Fig. 8 for the parameter sets  $iii$, $p$ and $m^*$. One
cannot draw a firm conclusion as to which set is better. It
should be noted, however, that the predictions of $m^*$
for the $B(E2)$ transitions and the charge radii seem to be
much closer to the empirical values than those for the other
forces. 

In Fig. 9 the neutron skin thickness ($r_n$-$r_p$) is
plotted against mass number $A$ for the same three Skyrme
sets. It is seen that for the iii set the smallest neutron skin
thickness appears. The other two sets give similar results of the
difference in radii not exceeding 0.15 fm.  Finally in Fig.
10 the differences ($\beta_n$-$\beta_p$) of the neutron and
proton deformation parameters are shown. It is seen that
the difference increases with the mass number. It
is also seen that for heavier nuclei the proton
distribution appears to be more deformed than the neutron
distribution  as was already noticed in Refs. \cite{Ba95,Ba96}.
 
A similar procedure was also followed for the RMF theory.
In this case the parameter sets NL--1, NL--3, NL--SH  (see
Table 3) were employed.

In Fig. 11 the mean square deviations for the various observables
calculated with the three parameter sets are shown. It is
seen that all sets give satisfactory results with NL--3 and
NL--SH being slightly better.  In Fig. 12 the differences of
the calculated values from the experimental ones are given
as functions of the mass number. Especially
the masses are reproduced much better with NL--3.  In Figs.
13 and 14 the neutron skin and the difference of the
proton neutron deformation parameters are plotted against
mass number $A$.  The behaviour is rather similar in all
parametrizations.  For NL--1 the neutron skin is larger than
in the other two forces.  This could be attributed to the
large asymmetry energy of NL--1.  The differences of the
proton neutron deformation parameters show rather similar
behaviour in all sets \cite{Ba95,Ba96}.

\newpage

\section{Conclusions}

In this work a systematic study of the ground state
properties of $\beta$ stable nuclei has been performed
within various effective microscopic theories and potential
models.   Best parametrizations among the most frequently
used have  been employed for each theoretical approach.
The aim was an investigation of the behaviour of the
various parameter sets over a wide range of mass numbers.

The best estimates of the mean square charge radii for the
$\beta$-stable nuclei are obtained with the Nilsson
potential with the parameters taken from Ref. \cite{Se86}.
It turned out that  the Chepurnov \cite{Chep} parametrization gives
on the average the best results among the considered
parameter sets of  the Saxon-Woods potential.

For Skyrme theory the effective forces $m^*$ \cite{Ba82} and $p$
\cite{Do83} seem to give the most satisfactory results for the
calculated observables.

In RMF theory the lagrangian parametrizations NL--1, NL--3 
and NL--SH have been considered. The study showed that along
$\beta$ stability line all sets work well with the NL--3 \cite{La96}
force being somewhat better, especially in the binding
energy predictions.

A few more comments are in place:
\begin{enumerate}

\item In the microscopic theories (Skyrme or RMF) the
neutron skin for the $\beta$--stable nuclei grows with A.
In the Skyrme mean field the neutron mean square radii are
larger than the proton ones by 0.15~fm for the heaviest
isotopes. In the RMF theory the neutron skin is somewhat
larger, being about 0.25~fm.

\item Both Skyrme mean field and RMF present a significant
difference in the quadrupole deformation of the neutron and
proton distributions, the protons distribution being more
deformed than the neutrons one.  This difference can
become larger than 10\% of the total magnitude of the
deformation.

\item These two effects should be taken into account in the
macroscopic-microscopic models, e.g. in the liquid droplet
model plus the Strutinsky shell correction for the
Saxon-Woods or Nilsson single particle levels. The effect
of the different deformations of the proton and neutron
distribution can be also important for the calculation of
the fission barriers using potential models. This was already
suggested in Ref. \cite{Ba95,Ba96}.
\end{enumerate}

The different density distributions of protons and
neutrons in a nucleus should be taken into account in all
the calculations where the collective variables enter
parametrically in order to minimize the potential energy of the
nucleus. We think that it would be worthwhile to look 
at proton and neutron equilibrium deformation separately in the
macroscopic-microscopic type of calculations. 

\newpage
{\bf Acknowledgements}

\bigskip\noindent
Krzysztof Pomorski gratefully acknowledges the warm
hospitality extended to him by the Theoretical Physics
Group of the Technical University in M\"unchen as well as a
grant from the Deutsche Forschung Gemeinschaft. Another
author (G.A.L) acknowledges the warm hospitality of the
theory group of the Maria Curie Sk{\l}odowska University during his 
short stay in Lublin, and  financial support by the E.U. under
the contract TMB-EU/ERB FMBCICT-950216. The authors thank
Professor Klaus Dietrich of TUM for helpful discussions.

\newpage
\section{Table captions}

\begin{enumerate}
\item Different sets of parameters of the Saxon-Woods potential
(Eq. \req{ws}). "universal" \cite{Cw87}, Wahlborn
\cite{Wahl}, Rost \cite{Rost}, Chepurnov \cite{Chep} and the "new"
\cite{New} ones were chosen.

\item Parameters of the Skyrme forces (Eq. \req{Skyrme})
used in our calculations. 
$i$ \cite{Va72}, $ii$ \cite{Va73}, $iii$, $iv$, $v$, $vi$
\cite{Be75}, $vii$ \cite{Gi80}, $m^\star$ \cite{Ba82}, $a$ \cite{Ko76}, 
$p$ \cite{Do83}.

\item Parameters used in the relativistic  mean--field lagrangian, Eq. \req{lrmf}.
 NL-1 \cite{Re86}, NL-SH \cite{Sh93}, NL-3 \cite{La96}.
\end{enumerate}


\section{Figure captions}
\begin{enumerate}
\item The estimated {\it 'experimental'} position of the proton 
      Fermi levels (diamonds, 'exp.') for even--even nuclei along the 
      $\beta$--stability line. The solid line (avr.) was obtained from the
      phenomenological formula \req{avlp}--\req{avln}.
\item The same as in Fig.1 but for neutrons.
\item The charge root mean square radii obtained with the
      Nilsson potential (solid line, 'Nil.') are compared with the
      experimental data \cite{Vr87,Ot89} (diamonds, 'exp.').
\item The reduced transition probabilities B(E2) obtained with the
      Nilsson potential (solid line, 'Nil.') compared with the
      experimental data (diamonds, 'exp.') \cite{Ot89}.
\item The differences between the theoretical and experimental \cite{Au93}
      separation energies of protons $S_p$ and neutrons $S_n$. The
      theoretical estimates 
      were obtained with the Saxon-Woods potential with the Chepurnov
      \cite{Chep} parameters set (solid line, 'Chep.') and universal set 
      \cite{Cw87} (dashed line, 'Univ.').
\item The root mean square errors for the neutron ($S_n$) and proton ($S_p$)
      separation energies, the reduced transition probabilities $B(E2)$
      and the charge radii ($r_{ch}$) obtained within different 
      Saxon-Woods potentials (Table 1).
\item The root mean square errors for the neutron ($S_n$) and proton ($S_p$)
      separation energies, the binding energies ($B$), the reduced transition 
      probabilities  $B(E2)$ and charge radii ($r_{ch}$) obtained within 
      the HFB calculation for different sets of the Skyrme forces 
      (Table 2).
\item The differences between the theoretical and experimental \cite{Au93}
      separation energies of neutrons ($S_n$) and protons ($S_p$), the 
      binding energies ($B$), the reduced transition probabilities $B(E2)$
      and the charge radii ($r_{ch}$) for the nuclei along the $\beta$--stability line. 
      The theoretical estimates were obtained within the HFB calculation
      for $iii$, $p$ and $m^*$ sets of the Skyrme forces (Table 2).
\item The differences (diamonds) between the mean square radii of the neutron 
     ($r_n$) and proton ($r_p$) density distributions evaluated within the HFB 
      calculation with the $iii$, $p$ and $m^*$ Skyrme forces (Table 2) for 
      the nuclei along the $\beta$--stability line.
\item The difference (diamonds) between the quadrupole deformation of the neutron
      ($\beta_n$) and proton ($\beta_p$) density distributions obtained 
      within the HFB calculation with the $iii$, $p$ and $m^*$ Skyrme 
      forces  (Table 2) for the nuclei along the $\beta$--stability line.
\item The same as in Fig. 7 but for the estimated obtained within the RMF 
      theory for NL-3, NL-SH and NL-1 sets of parameters (Table 3).
\item The same as in Fig. 8 but for the estimates obtained within the RMF 
theory for NL--3, NL--SH and NL--1 sets of parameters (Table 3).
\item The same as in Fig. 9 but for the estimates obtained within the RMF 
theory for NL--3, NL--SH and NL--1 sets of parameters (Table 3).
\item The same as in Fig. 10 but for the estimates obtained within the RMF 
theory for NL--3, NL--SH and NL--1 sets of parameters (Table 3).
      
\end{enumerate}

\newpage

\begin{center}
{\bf Table 1}\\
\vspace{1cm}
\begin{tabular}{|c|c|c|c|c|c|c|}
\hline\hline
            &     &           &           &        &          &        \\
  Parameter &units& Universal & Wahlborn  &  Rost  & Chepurnov&  New   \\
            &     &           &           &        &          &        \\
\hline      
            &     &           &           &        &          &        \\
$ V_0 $     & MeV &  49.6     &   51.0    & 49.6   & 53.3     &  49.6  \\ 
            &     &           &           &        &          &        \\
$ \kappa   $& -   &   0.86    &    0.67   &  0.86  &  0.63    &   0.86 \\
            &     &           &           &        &          &        \\
$ a   $     & fm  &   0.70    &    0.67   &  0.70  &  0.63    &   0.70 \\
            &     &           &           &        &          &        \\
$ r_n $     & fm  &   1.347   &    1.27   &  1.347 &  1.24    &   1.347\\
            &     &           &           &        &          &        \\
$ \lambda_n$& -   &  35.0     &   32.0    & 31.5   &23.8$\cdot$(1+2I)& * \\
            &     &           &           &        &          &        \\
$ r_n^{so} $&  fm &   1.31    &    1.27   &  1.280 &  1.24    &    *   \\
            &     &           &           &        &          &        \\
$ r_p $     &  fm &   1.275   &     1.27  &  1.275 &  1.24    &   1.275\\
            &     &           &           &        &          &        \\
$ \lambda_p$& -   &  36.0     &    32.0   &  17.8  &23.8$\cdot$(1+2I)& * \\
            &     &           &           &        &          &        \\
$ r_p^{so} $&  fm &   1.32    &     1.27  &   0.932&  1.24    &    *   \\
            &     &           &           &        &          &        \\
\hline
\end{tabular}
\end{center}  

$^*${ \footnotesize The radius constant and the strenght of the spin-orbit 
 potential  are deformation dependent as described in Ref. \cite{New}.}

\newpage

\begin{center}
                               {\bf Table 2}\\
\vspace{1cm}
\begin{tabular}{|c|c|c|c|c|c|c|}
\hline\hline
&&&&&&\\
Parameter & unit   &   $i$    & $ii$    & $iii$   & $iv$    & $v$ \\
&&&&&&\\
\hline 
$t_0 $& MeV fm$^3$ & -1057.30 &-1057.30 &-1128.75 &-1205.60 &-1248.29
\\
$t_1 $& MeV fm$^5$ & 235.90   &235.90   &395.00   &765.00   &970.56 \\
$t_2 $& MeV fm$^5$ & -100.00  &-100.00  &-95.00   &35.00    &107.22 \\
$t_3 $& MeV fm$^{(3+3\gamma)} $
                   & 14463.5  &14463.5  &14000.0  &5000.0   &0.0 \\
$x_0 $& -          & 0.560    &0.560    &0.450    &0.050    &-0.170 \\
$x_1 $& -          & 0        &0        &0        &0        &0 \\
$x_2 $& -          & 0        &0        &0        &0        &0 \\
$x_3 $& -          & 0        &0        &0        &0        &0 \\ 
$\gamma $& -       & 1        &1        &1        &1        &1 \\
$w_0 $& MeV fm$^5 $& 120.0    &105.0    &120.0    &150.0    &150.0 \\
\hline
&&&&&&\\
Parameter & unit   & $vi$     & $vii$   & $p$      & $m^*$   & $a$ \\
&&&&&&\\
\hline
$t_0 $& MeV fm$^3$ & -1101.81 &-1096.00 &-2931.70 &-2645.00 &-1602.78
\\
$t_1 $& MeV fm$^5$ & 271.67   &246.20   &320.62   &410.00   &570.88 \\
$t_2 $& MeV fm$^5$ & -138.33  &-148.00  &-337.41  &-135.00  &-67.70 \\
$t_3 $& MeV fm$^{(3+3\gamma)} $
                   & 17000.0  &17626.0  &18708.97 &15595.0  &8000.0 \\
$x_0 $& -          & 0.583    &0.620    &0.29215  &0.090    &-0.020 \\
$x_1 $& -          & 0        &0        &0.65318  &0        &0 \\
$x_2 $& -          & 0        &0        &-0.53732 &0        &0 \\  
$x_3 $& -          & 0        &1        &0.18103  &0        &0.286 \\
$\gamma $& -       & 1        &1        & 1/6     & 1/6     & 1/3 \\
$w_0 $& MeV fm$^5 $& 115.0    &112.0    &100.0    &130.0    &125.0 \\
\hline
\end{tabular}
\end{center}

%

\newpage

\begin{center}
{\bf Table 3}\\
\vspace{1cm}
\begin{tabular}{|c|c|c|c|c|}
\hline\hline
           &     &          &          &             \\
Parameter  &units&  NL--1     &  NL--SH   &  NL--3        \\
           &     &          &          &             \\
\hline
           &     &          &          &             \\
$ M       $&MeV  & 938.0    & 939.0    & 939.0       \\ 
$ m_\sigma$&MeV  & 492.25   & 526.059  & 508.194     \\
$ m_\omega$&MeV  & 795.359  & 783.00   & 782.501     \\
$ m_\rho  $&MeV  & 763.000  & 763.00   & 763.000     \\
$ g_\sigma$& -   &  10.138  &  10.4444 &  10.217      \\
$ g_2     $&$fm^{-1}$& -12.172  &  -6.9099 & -10.431     \\
$ g_3     $& -   & -36.265  & -15.8337 & -28.885     \\
$ g_\omega$& -   &  13.285  &  12.945  &  12.868     \\
$ g_\rho  $& -   &   4.976  &   4.383  &   4.474     \\
           &     &          &          &             \\
\hline
\end{tabular}
\end{center}

\newpage

\ifig{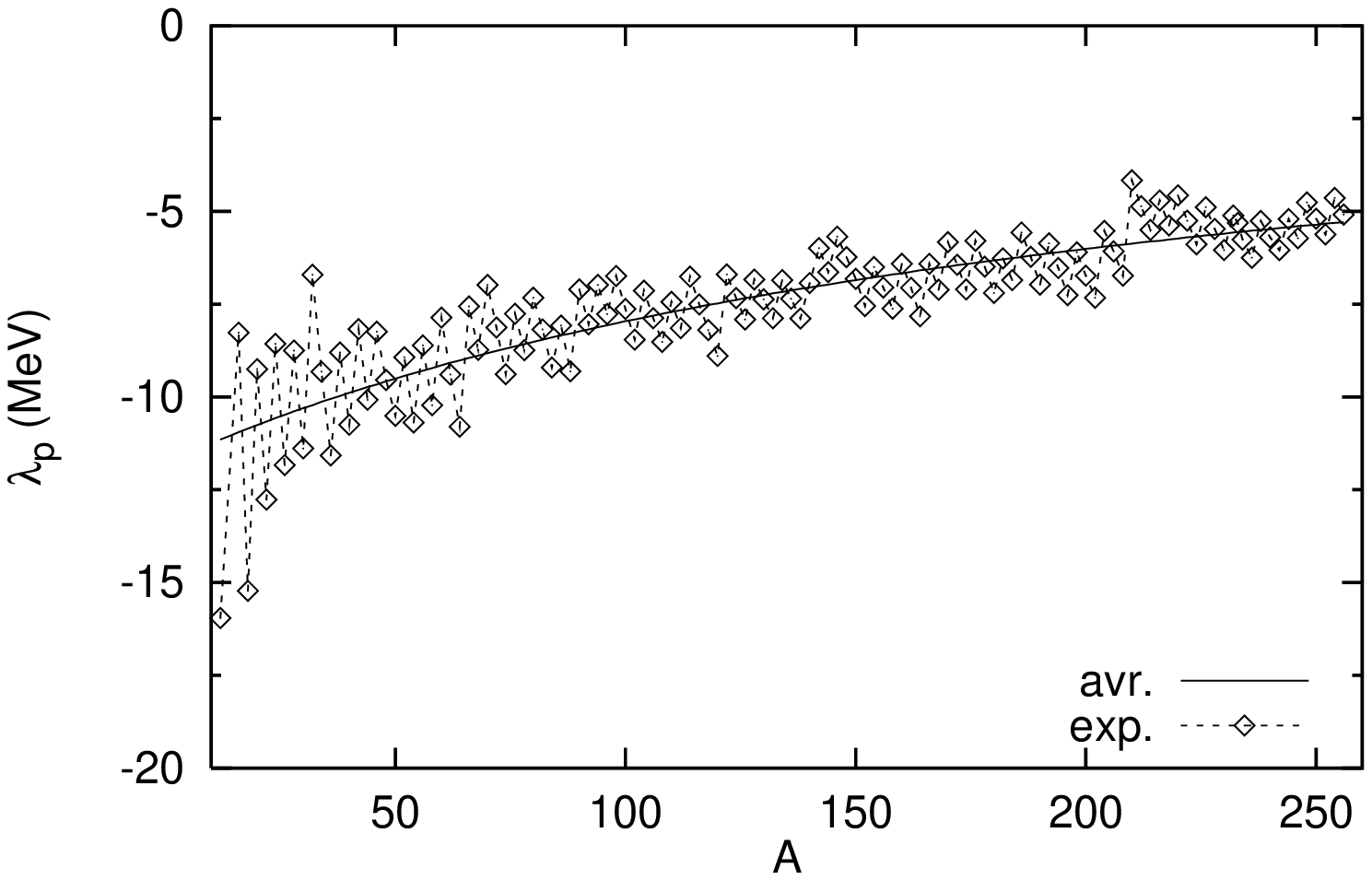}{60mm}{1}{} 
\ifig{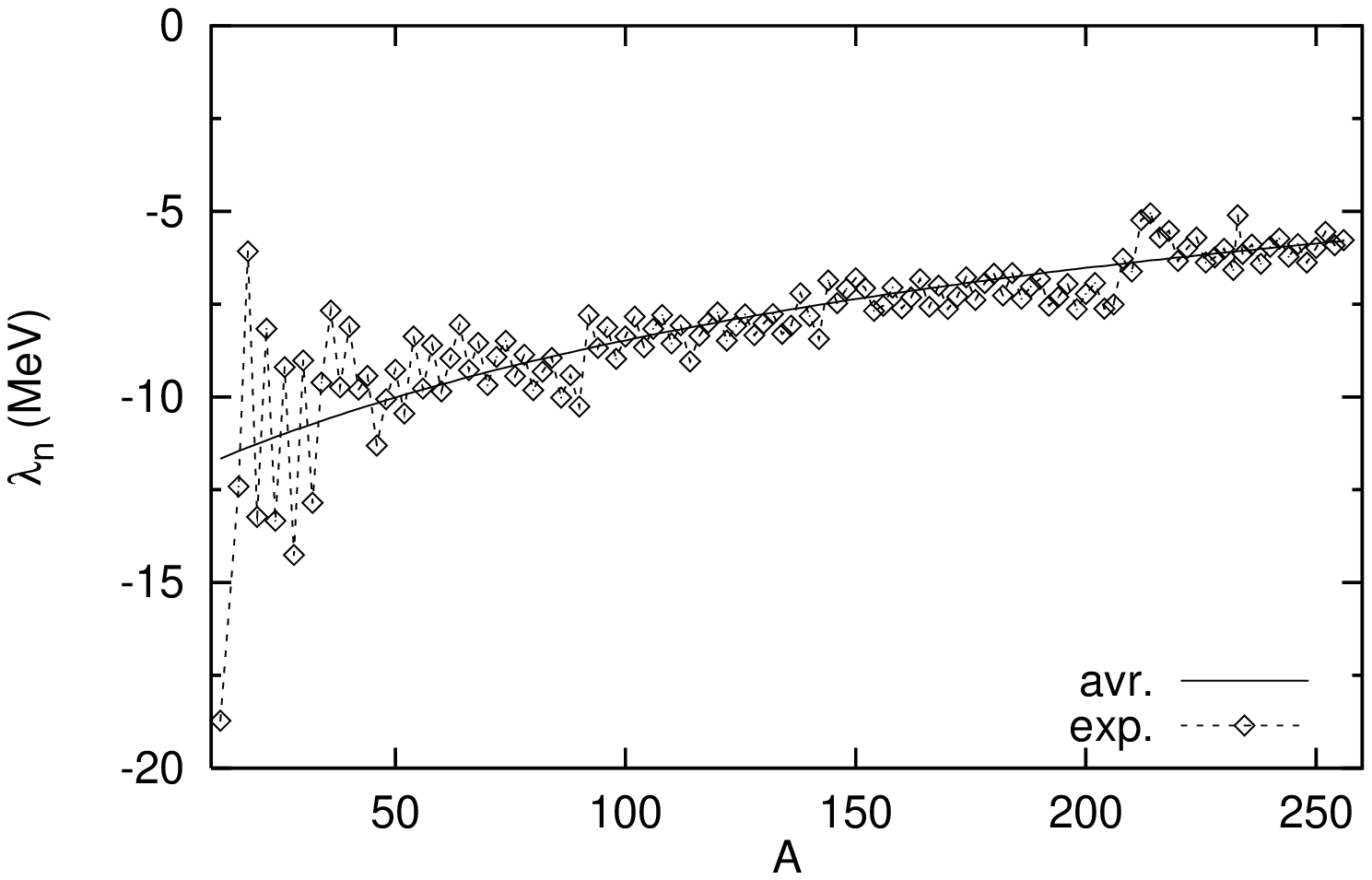}{60mm}{2}{}
\pagebreak[5]
 
\ifig{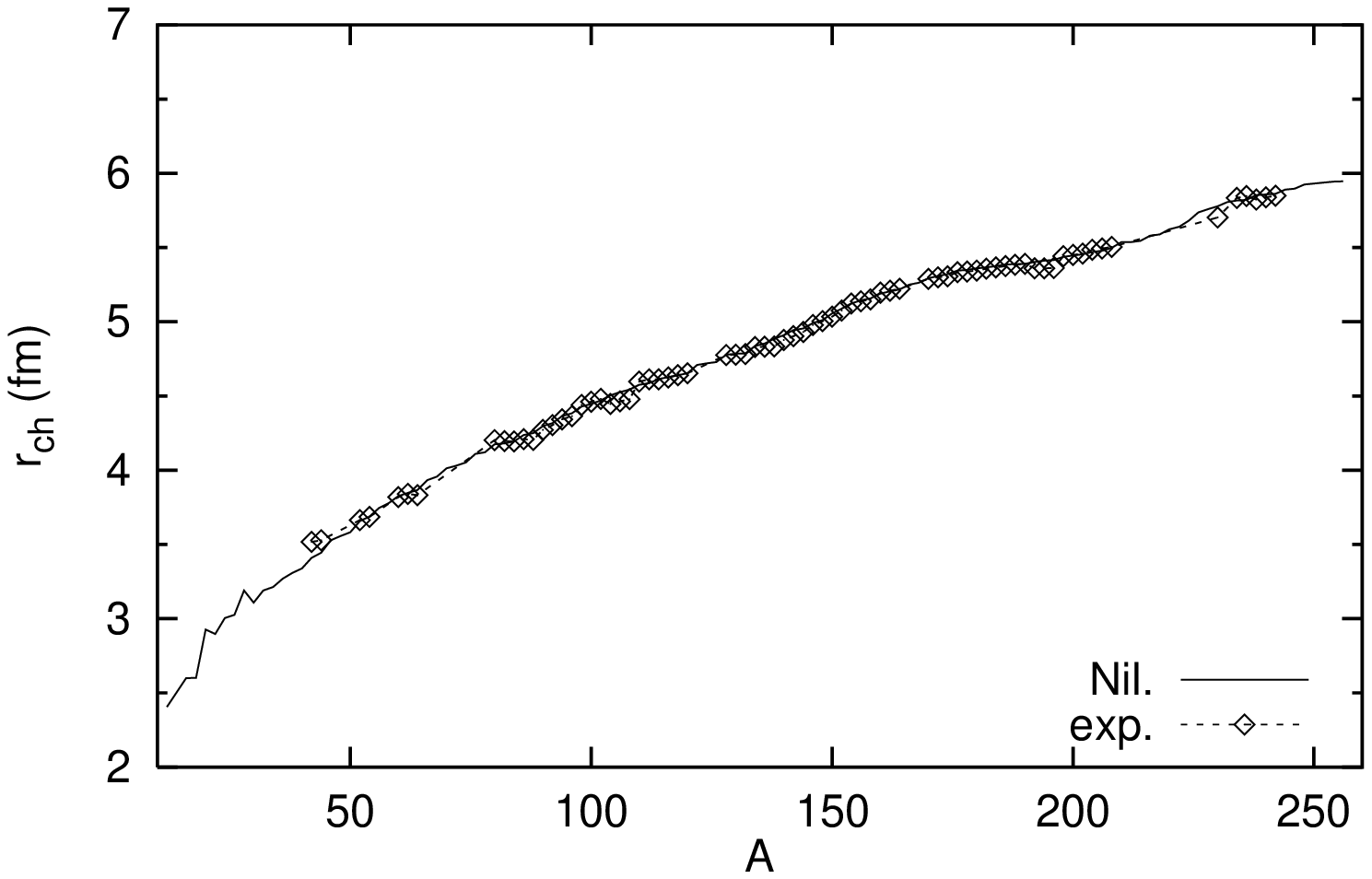}{60mm}{3}{} 
\ifig{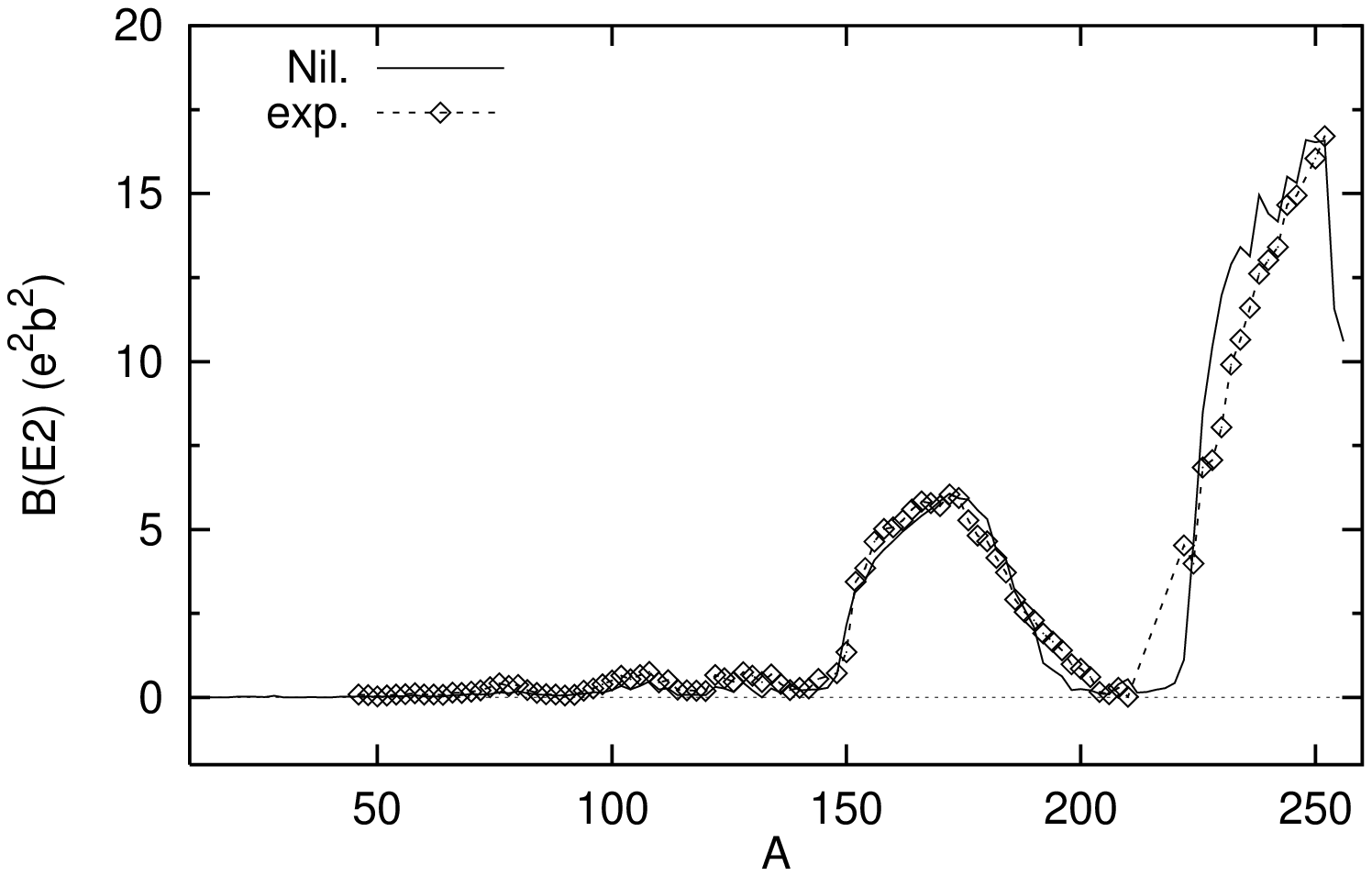}{60mm}{4}{} 
\pagebreak[5]

\ifig{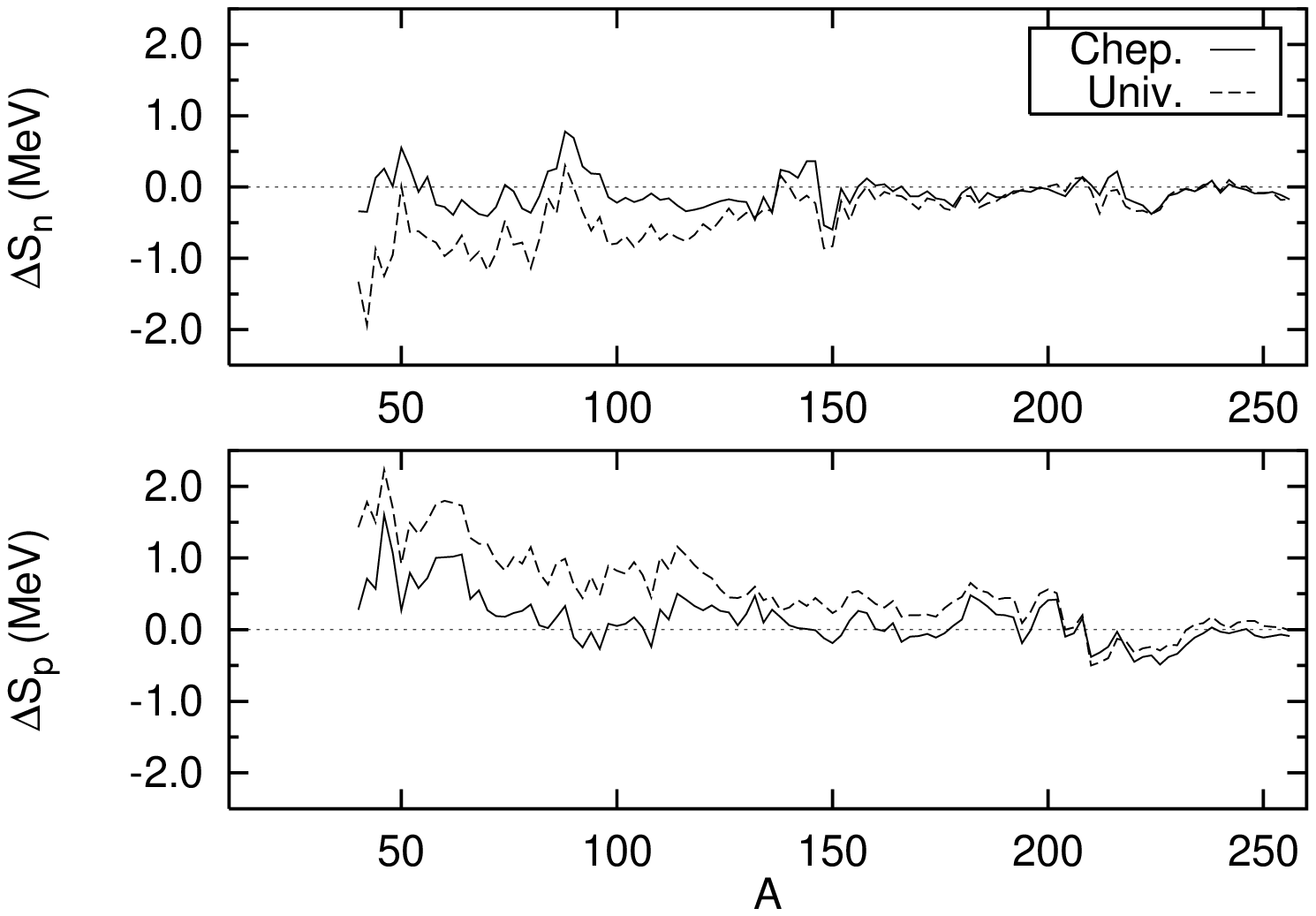}{100mm}{5}{}

\pagebreak[5]

\ifig{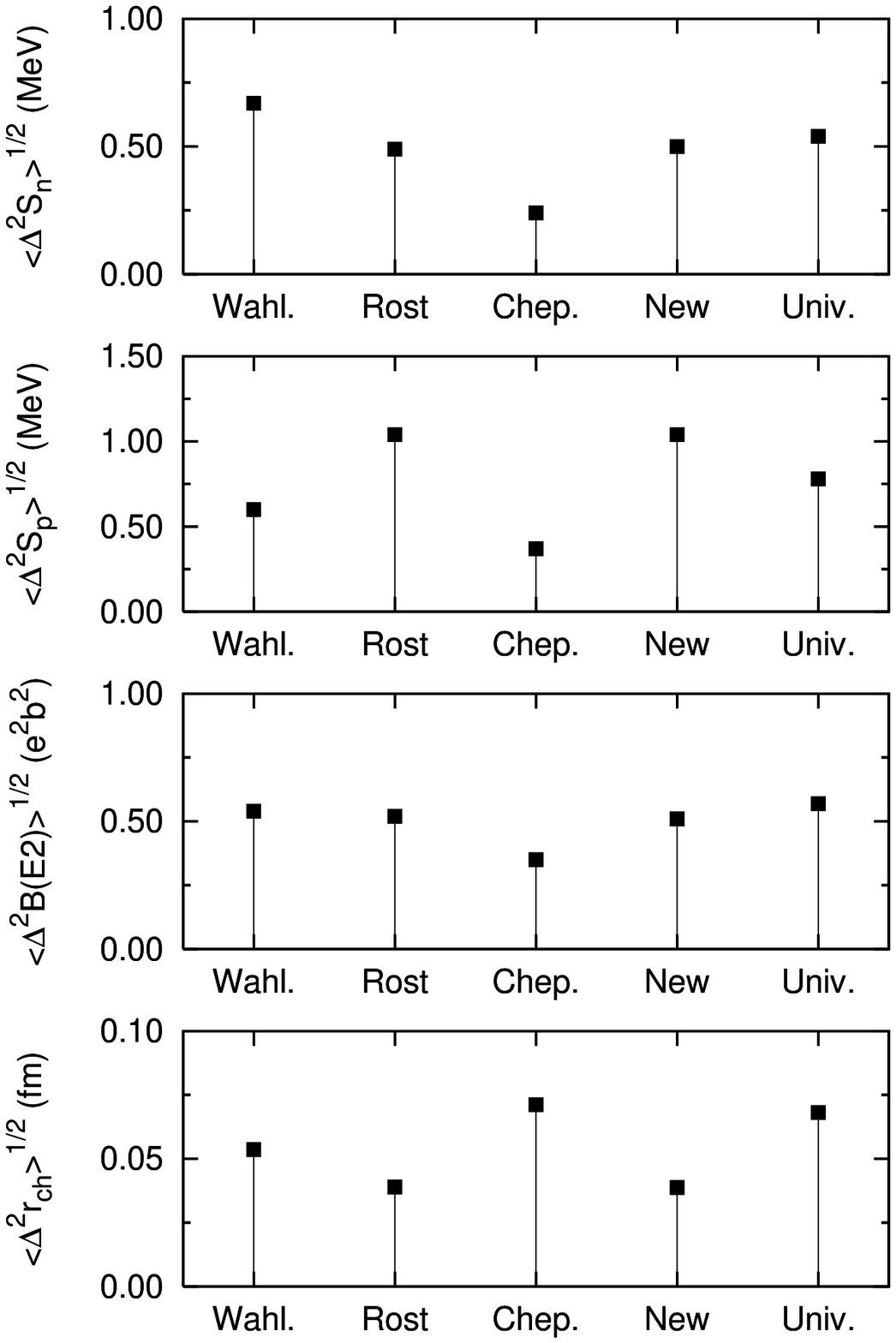}{160mm}{7}{}

\newpage

\ifig{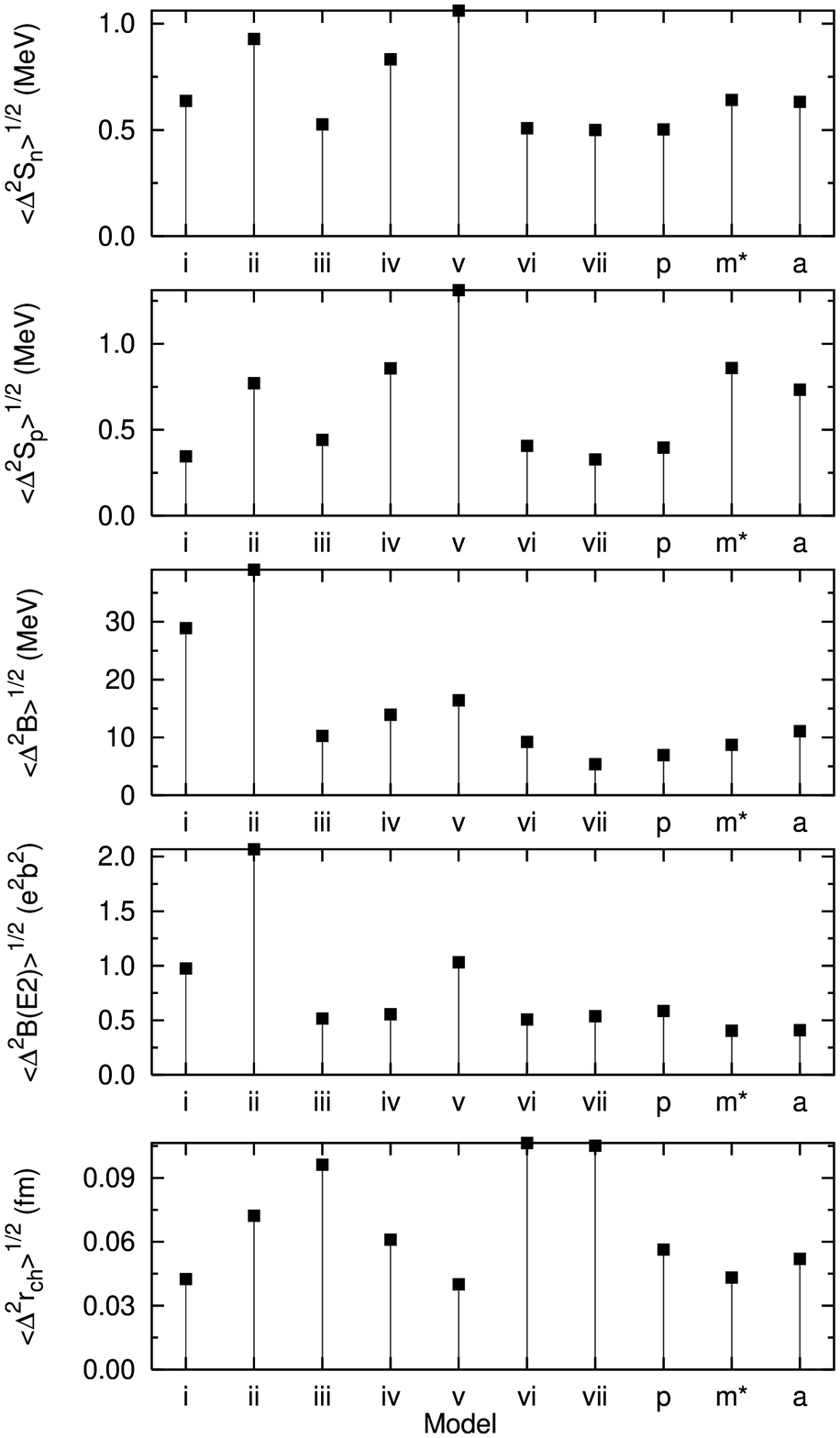}{160mm}{8}{}
\newpage

\ifig{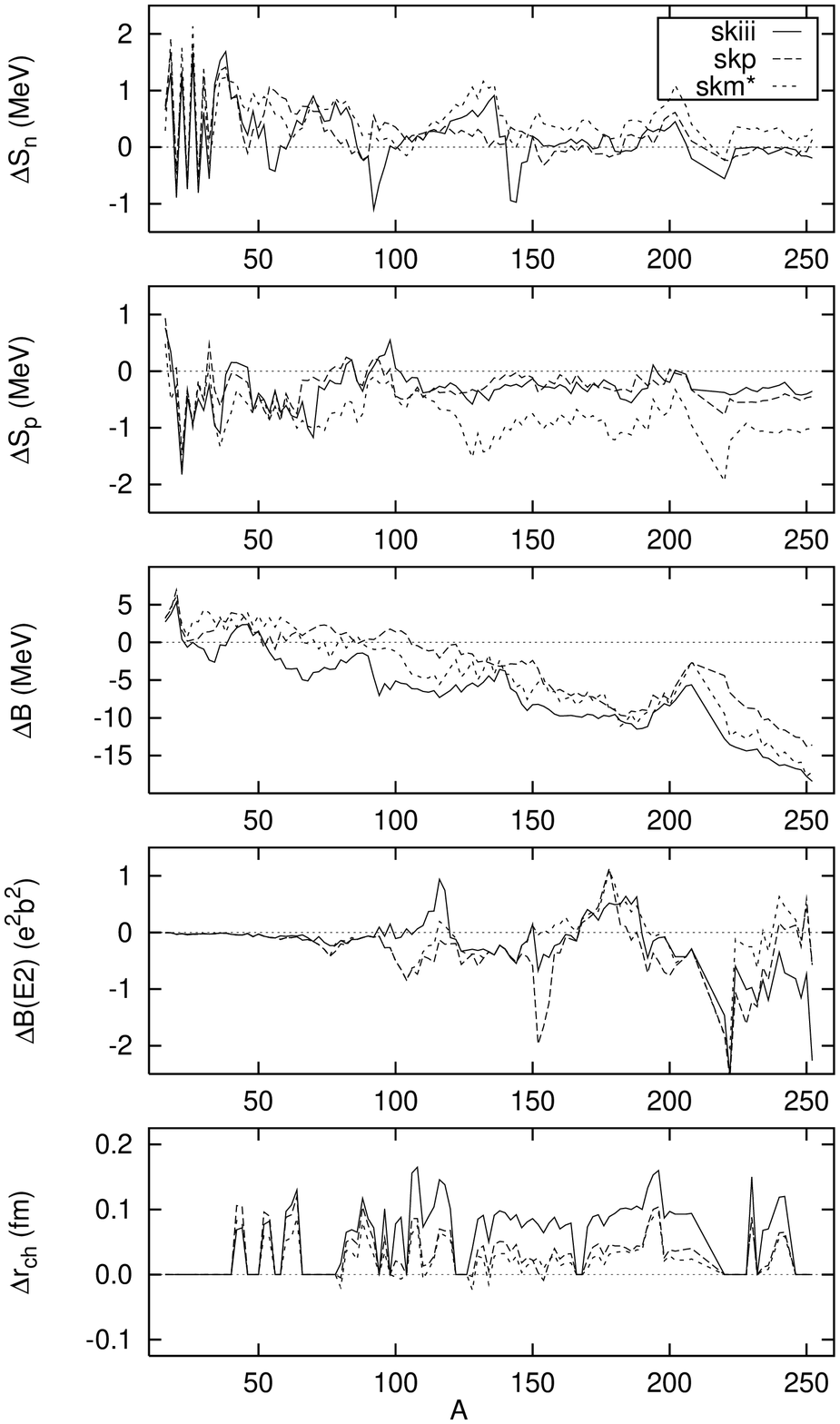}{160mm}{9}{}
\newpage

\ifig{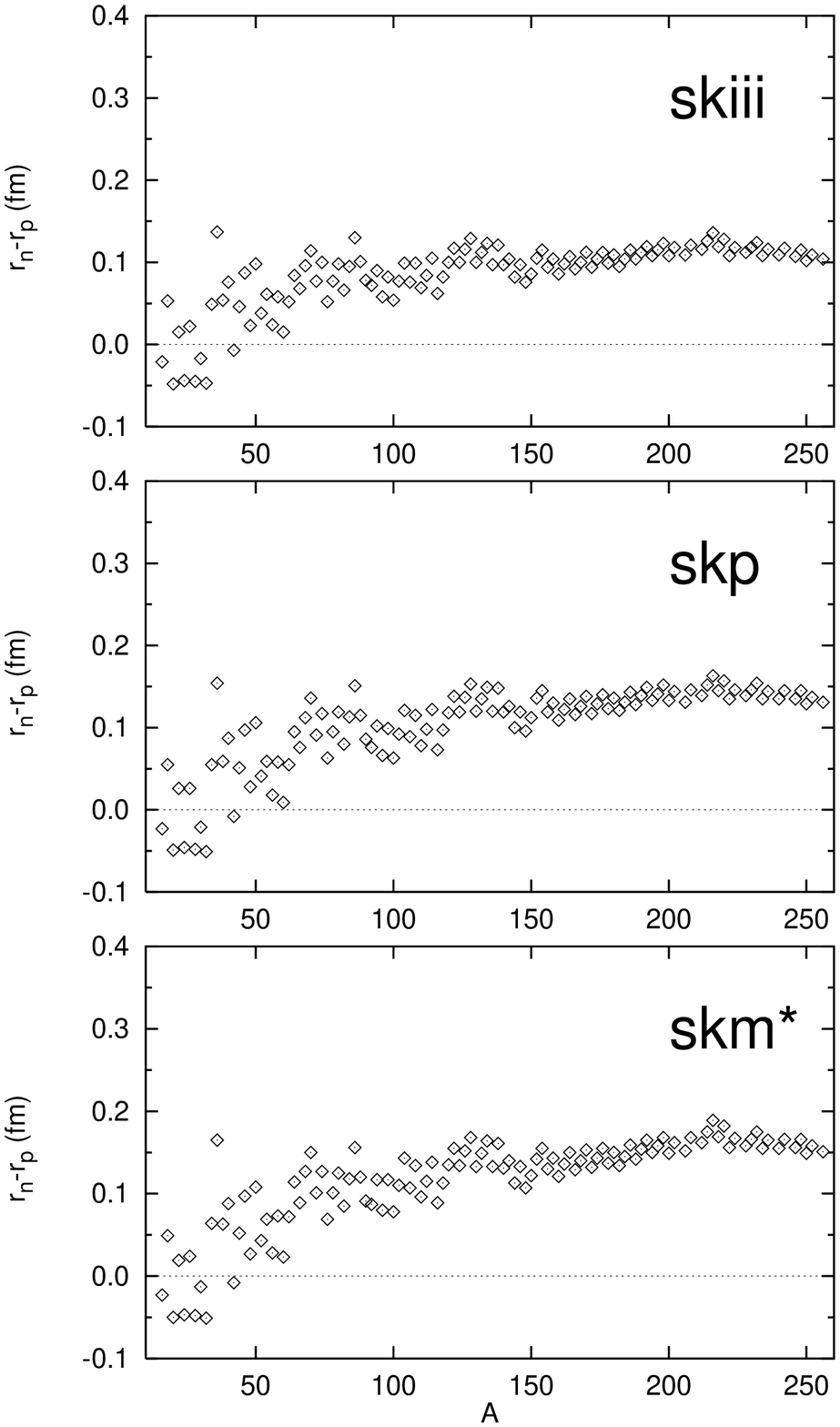}{160mm}{10}{}
\newpage

\ifig{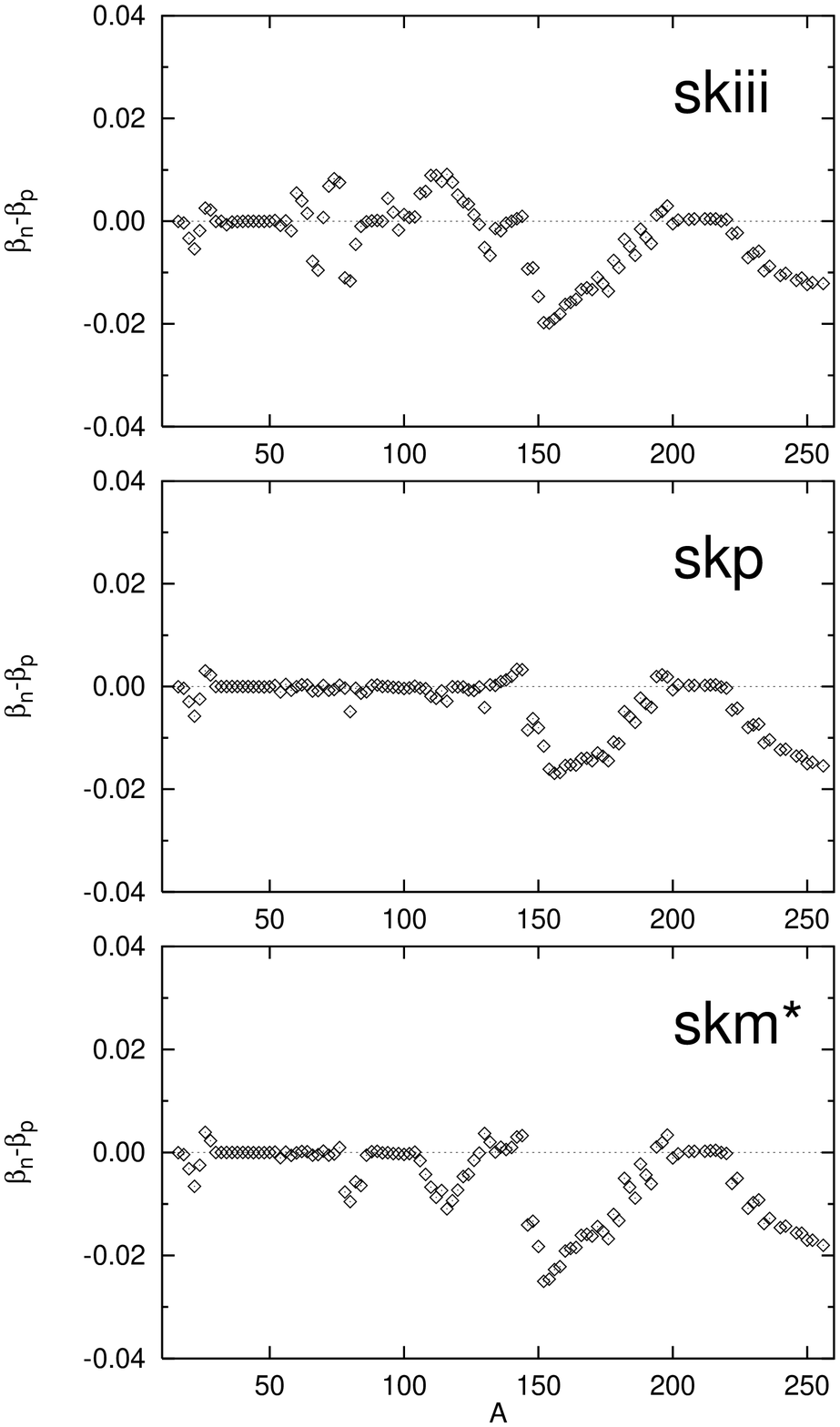}{160mm}{11}{}
\newpage

\ifig{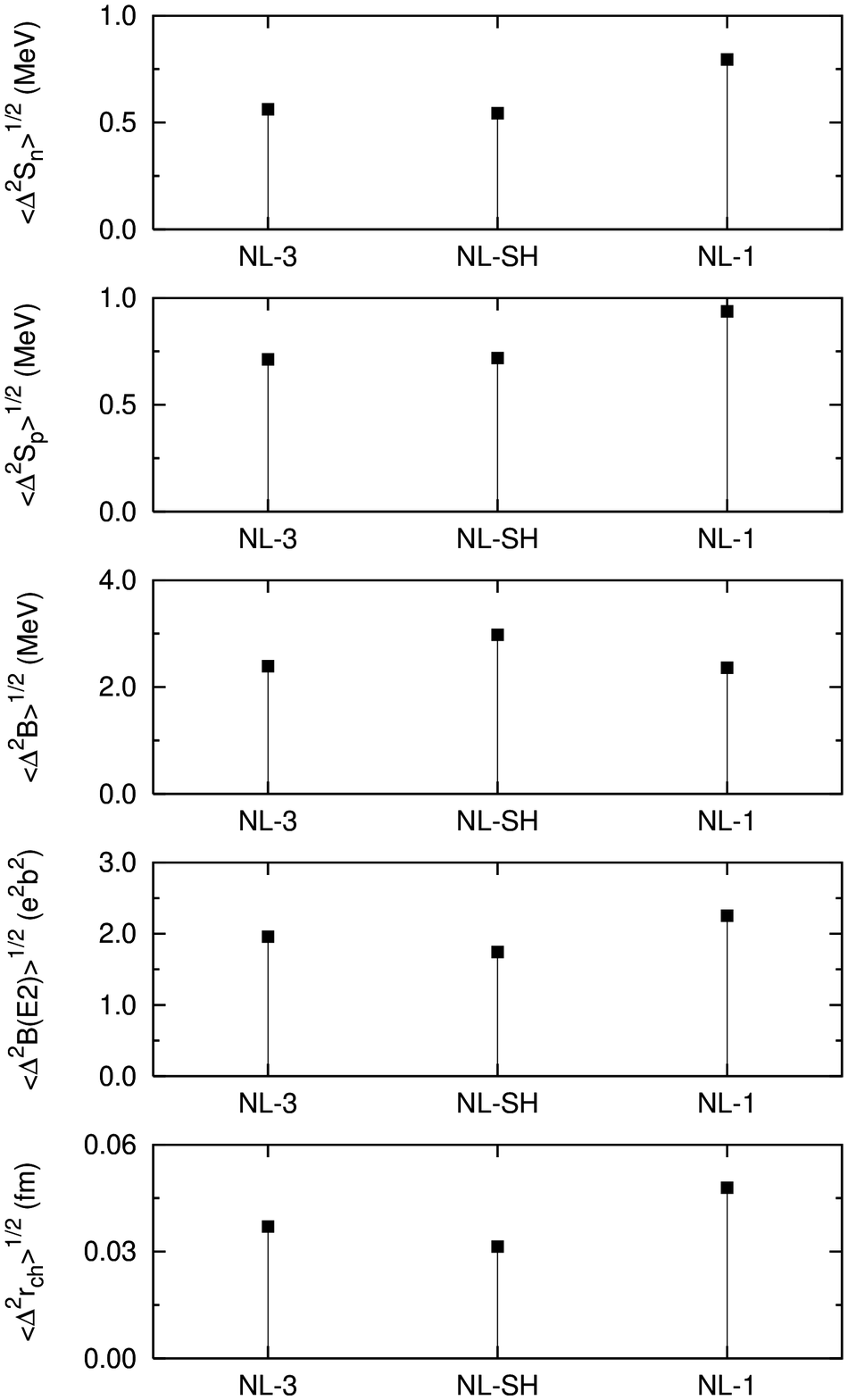}{160mm}{12}{} 

\newpage

\ifig{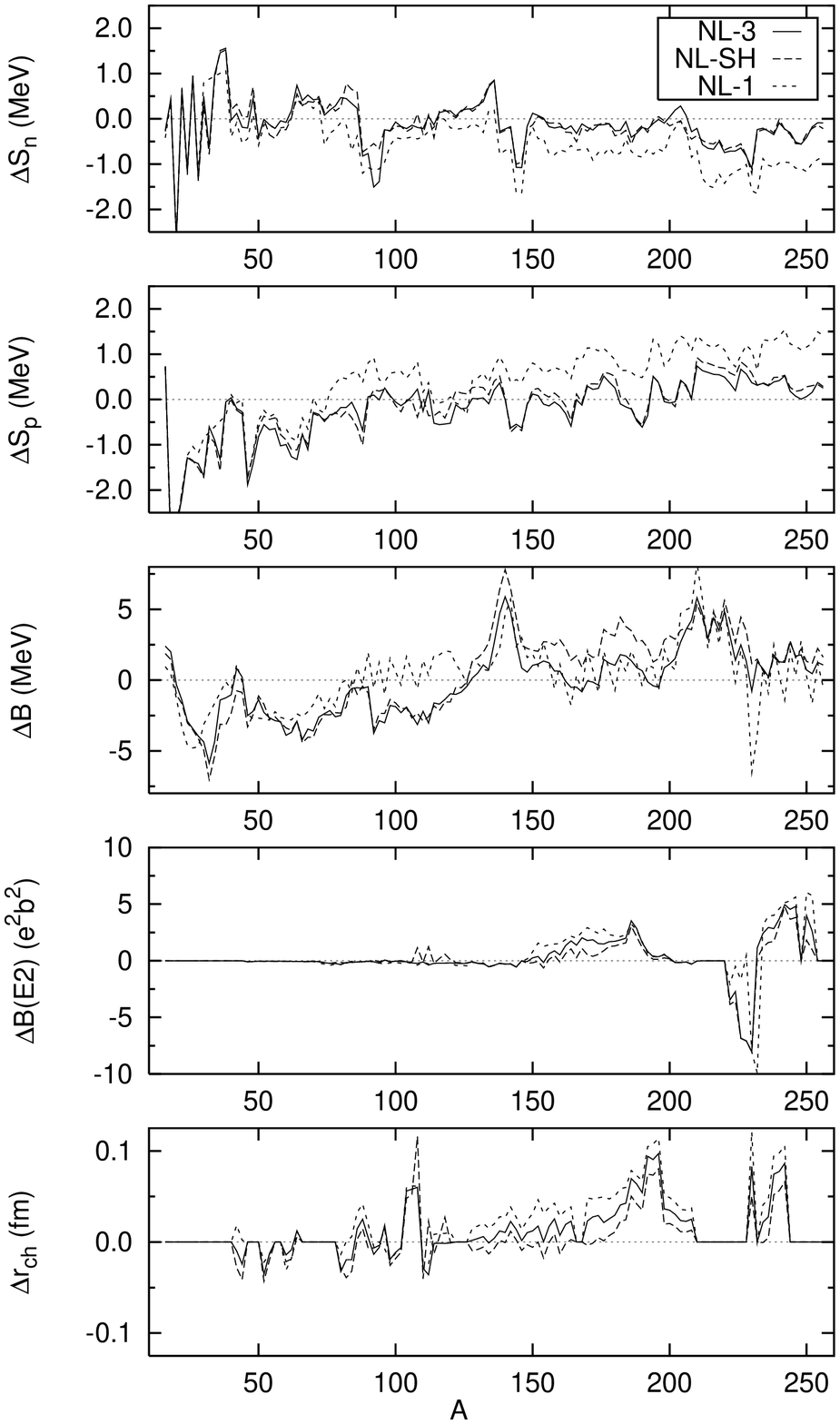}{160mm}{13}{} 
\newpage

\ifig{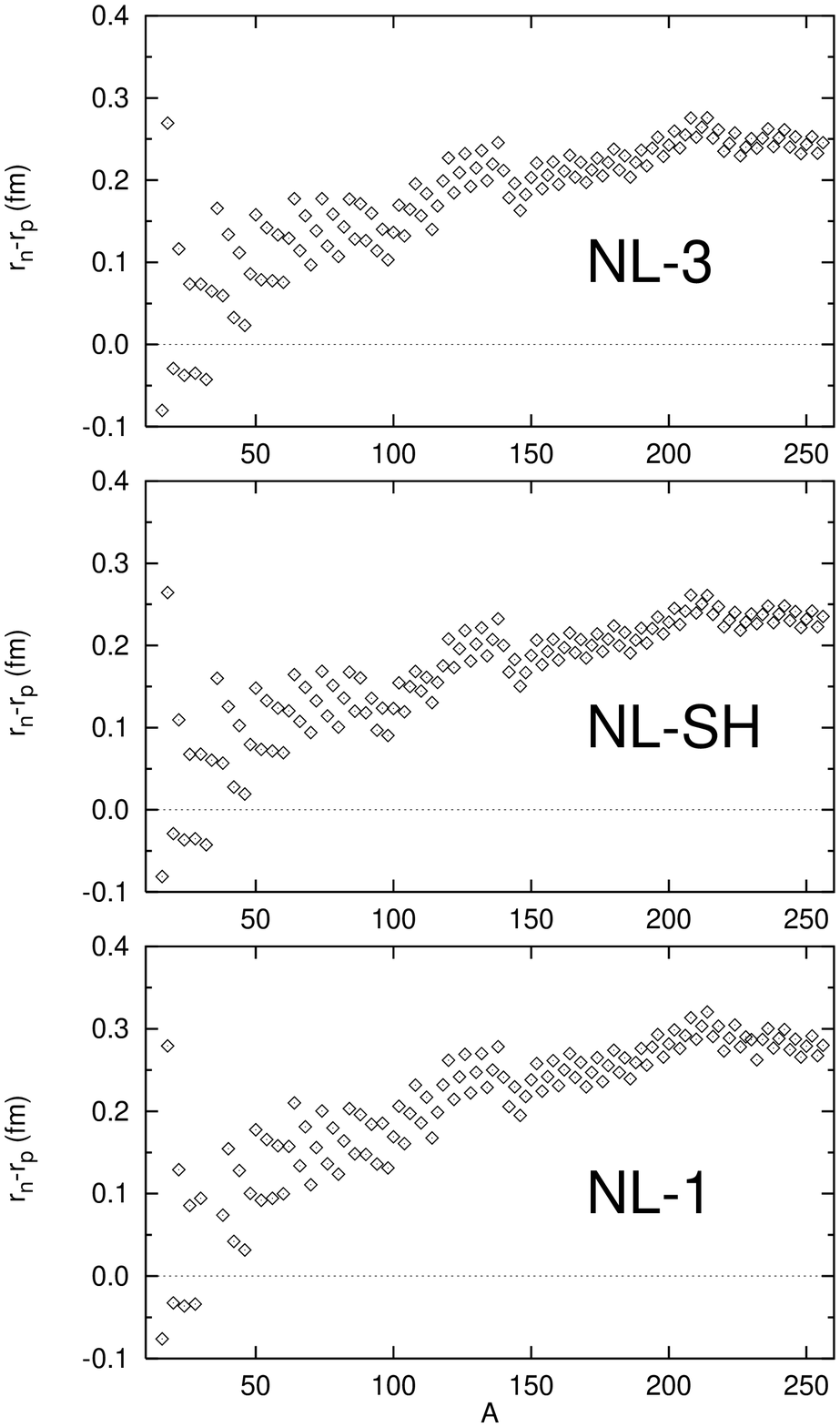}{160mm}{14}{} 
\newpage

\ifig{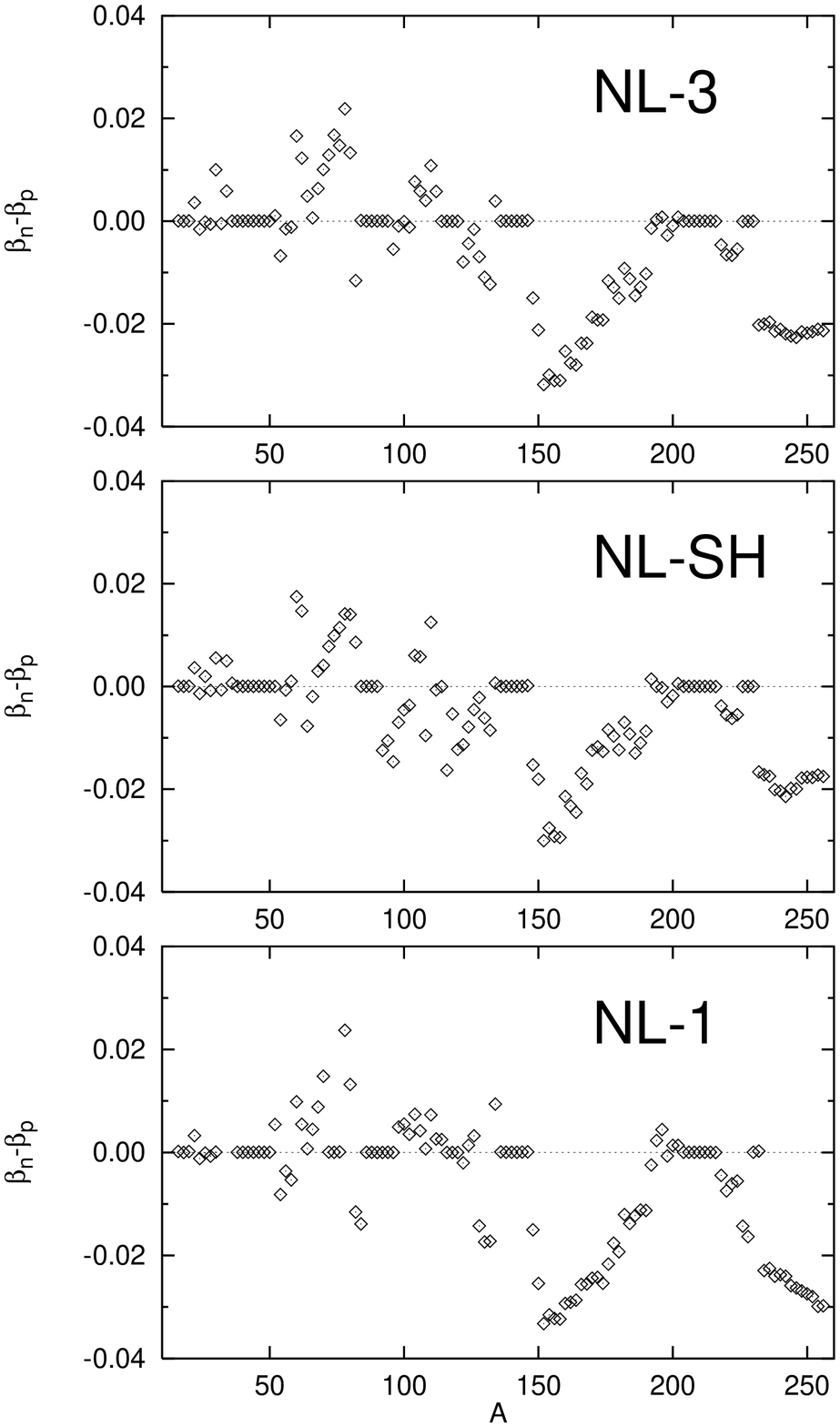}{160mm}{15}{}

\end{document}